\documentclass[aps, pra, amsmath, superscriptaddress,preprintnumbers,twocolumn]{revtex4-2}
\usepackage{graphicx}
\usepackage{dcolumn}
\usepackage{bm}
\usepackage{amsmath}
\usepackage{subfigure}
\usepackage{amsfonts}
\usepackage[svgnames]{xcolor}
\usepackage{longtable}
\usepackage{appendix}
\usepackage{multirow}
\usepackage{makecell}
\usepackage{booktabs}
\usepackage{tabularx}
\usepackage[colorlinks,
            linkcolor=blue,
			filecolor=blue,
			urlcolor= blue,
			citecolor=blue,
            ]{hyperref}
\usepackage{amsthm}
\usepackage{setspace}
\usepackage[T1]{fontenc}
\usepackage{extarrows}

\usepackage[percent]{overpic}
\usepackage{dsfont}
\usepackage{physics}

\newcommand{\ii}{\mathrm{i}}
\newcommand{\proj}[1]{\lvert #1\rangle\!\langle #1\rvert}
\newcommand{\CSS}{\mathrm{CSS}}

\begin{document}

\title{Efficient preparation of spin-squeezed states with Grover's algorithm}

\author{Zhi-Yuan Zeng}
\affiliation{Department of Applied Physics, University of Science and Technology Beijing, Beijing 100083, China}

\author{Wen-Long Ma}
\email{wenlongma@semi.ac.cn}
\affiliation{State Key Laboratory of Semiconductor Physics and Chip Technologies, Institute of Semiconductors, Chinese Academy of Sciences, Beijing 100083, China}
\affiliation{Center of Materials Science and Opto-Electronic Technology, University of Chinese Academy of Sciences, Beijing 100049, China}

\begin{abstract}
We propose a scheme for preparing spin-squeezed states in an ensemble of $N$ spin-$1/2$ particles based on Grover's algorithm. With only a few oracles, the scheme can exactly prepare superpositions of a few neighboring Dicke states with the Wineland squeezing parameter scaling as $N^{-1}$. The scheme requires only global rotations and Dicke-selective phase shifts, thus avoiding the need to engineer a dedicated many-body Hamiltonian for the squeezing dynamics. We further discuss the feasibility of realizing the scheme in cavity QED systems, with the phase-matched Grover iterations realized by single-photon reflection from a dispersively coupled cavity. Full-sequence scattering calculations accounting for finite photon bandwidth and loss show that appreciable spin squeezing can be achieved with high fidelity for suitable system parameters.

\end{abstract}

\maketitle

\section{Introduction}
\label{sec:introduction}

Collective spin squeezing, arising from many-body correlations, can enhance sensitivity to small rotations and phases. Spin squeezing below the standard quantum limit (SQL) has been exploited in metrological applications such as atomic clocks \cite{louchetchauvet2010,pedrozo2020,robinson2024,yang2025clock,eckner2023,zaporski2025}, magnetometry \cite{sewell2012,muessel2014,duan2025,zheng2023}, and atom interferometry \cite{gross2010,greve2022,cassens2025}.
For an ensemble of $N$ effective spin-$1/2$ particles, the Wineland metrological squeezing parameter is defined as
\begin{equation}
    \xi_R^2 \equiv
    \frac{N(\Delta J_\perp)^2_{\min}}
    {|\langle\mathbf{J}\rangle|^2},
    \label{eq:wineland}
\end{equation}
where $\mathbf{J}=(J_x,J_y,J_z)$ denotes the collective spin, and $(\Delta J_\perp)^2_{\min}$ is the minimized spin variance among all axes perpendicular to the mean collective spin polarization $\langle\mathbf J\rangle$~\cite{wineland1992,wineland1994,ma2011,pezze2018}. For states with nonzero mean spin, $\xi_R^2<1$ indicates a metrological gain beyond the SQL. 

Established approaches to preparing spin-squeezed states include
squeezing transfer~\cite{hald1999,dantan2004,benmaimon2024},
coherent interactions~\cite{kitagawa1993,liu2011,leroux2010prl,braverman2019,huang2021,wu2025,barberena2024},
measurement-induced squeezing~\cite{kuzmich2000,cox2016,duan2025,barberena2024,vasilyev2013},
and engineered dissipation~\cite{dallatorre2013,gonzaleztudela2013,groszkowski2022,vasilyev2013,sundar2024,benmaimon2024}. Squeezing transfer relies on an externally prepared squeezed field, with its performance limited by the input squeezing and transfer efficiency \cite{huang2021,hald1999,dantan2004}. Among coherent approaches, rapid adiabatic passage can prepare Dicke-state superpositions with $N^{-1}$ Wineland squeezing scaling, but requires a specific nonlinear collective Hamiltonian and shaped driving fields \cite{carrasco2024}. Selective Dicke-level shifts combined with global rotations provide another coherent preparation route, with numerically fitted $N^{-0.843}$ scaling \cite{zhang2024}. Measurement and feedback can also achieve $N^{-1}$ scaling, with control conditioned on measurement outcomes \cite{thomsen2002}. Reservoir engineering relies on tailored collective dissipation to drive asymptotic convergence toward a squeezed steady state with $N^{-1}$ scaling \cite{groszkowski2022,dallatorre2013}. 

Here, we propose an efficient preparation method that first specifies the structure of the target spin-squeezed state and then constructs its preparation using Grover's algorithm, without requiring a dedicated many-body Hamiltonian for the squeezing dynamics. The protocol achieves $O(N^{-1})$ Wineland squeezing using only global rotations and Dicke-selective phase operations, without relying on squeezed-light input, measurement feedback, or dissipative relaxation.

Grover's algorithm originally developed for quantum search \cite{grover1997} has been used for quantum-state preparation \cite{azuma2000,johnsson2020,nagib2025prl,nagib2025pra}. Grover's algorithm can amplify selected components of an initial state while preserving their relative amplitudes, thereby preparing the desired superposition \cite{brassard2002}. Standard Grover iterations typically employ $\pi$-phase inversions, and for a general initial overlap, the target state cannot always be reached exactly after an integer number of iterations \cite{brassard2002}. Phase-matched variants replace these fixed $\pi$-phase inversions with suitably chosen non-$\pi$ phase rotations satisfying an appropriate phase-matching condition, allowing the amplification dynamics to be synchronized with an integer number of iterations and enabling exact preparation of the target state in an integer number of steps under ideal conditions \cite{hoyer2000,long2001}. This feature is well suited to the target-driven preparation strategy adopted here, in which the target-state structure is specified first and then realized through amplitude amplification.

\begin{figure}[!t]
  \centering
  \includegraphics[width=\columnwidth]{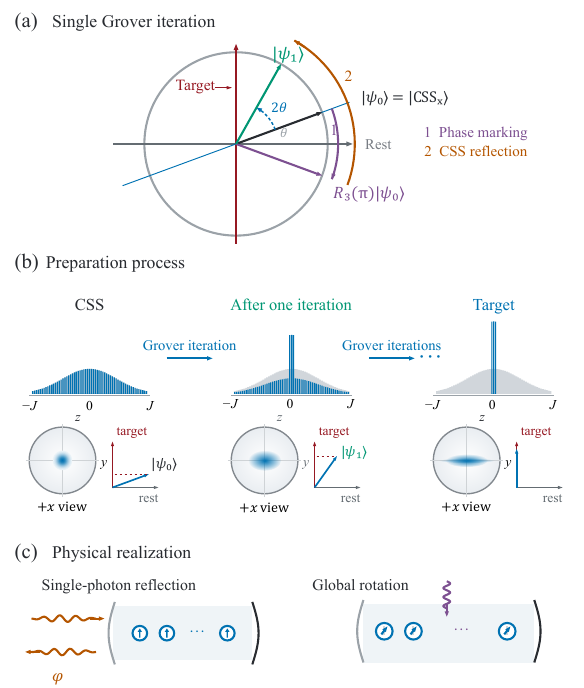}
\caption{
Grover-based preparation of a directly projected
three-component spin-squeezed state.
(a) Geometry of a standard Grover iteration with $\phi=\pi$,
where $R_3(\pi)\equiv\chi_{\mathrm{CSS}}(\pi)\chi_{\Pi_3}(\pi)$
denotes one Grover iteration for the three-component target subspace.
(b) Direct preparation by amplifying the $m=-1,0,+1$
components of an $x$-polarized coherent spin state.
Bars show schematic Dicke-amplitude magnitudes;
the $+x$ views illustrate transverse spin fluctuations.
(c) Physical building blocks for the cavity-QED implementation.
}
  \label{fig:preparation-overview}
\end{figure}

Previous studies have shown that superpositions of neighboring
Dicke states can exhibit metrologically useful spin squeezing
\cite{wineland1994,sorensen2001,rojo2003,huhu2014}.
We focus on symmetric superpositions of the $K$ neighboring
Dicke states closest to $m=0$, where $K$ denotes the number
of selected components.
The smallest nontrivial targets are the $m=\pm 1/2$
two-component state for odd $N\geq 3$ and the $m=-1,0,+1$
three-component state for even $N\geq 4$.
The odd-$N$ two-component target is the Yurke state,
a spin-squeezed state useful for enhanced phase sensitivity
in interferometry \cite{yurke1986,ono2017}.
Starting from an $x$-polarized coherent spin state,
we define the target state by retaining and normalizing
its amplitudes on the selected Dicke components.
Phase-matched Grover amplitude amplification prepares
these targets exactly in the ideal phase-oracle model
(Fig.~\ref{fig:preparation-overview}).
Previous cavity-QED proposals for Grover-based Dicke-state
preparation show that global rotations and Dicke-selective
phases induced by single-photon reflection can provide
the operations required for our scheme without a dedicated
squeezing Hamiltonian \cite{nagib2025prl,nagib2025pra}.

Our analysis shows that these targets are spin squeezed
and, for fixed $K$, exhibit $\xi_R^2\propto N^{-1}$
as $N$ increases.
The smallest targets yield $\xi_R^2\sim 4N^{-1}$
for odd $N$ ($K=2$) and $\xi_R^2\sim 6N^{-1}$
for even $N$ ($K=3$).
For even $N$, a single specially designed Grover iteration
first redistributes the relative amplitudes of the three
Dicke components, enabling subsequent phase-matched
Grover amplitude amplification to prepare a reshaped
target with the leading coefficient reduced from $6$
to $3$ at fixed mean-spin contrast.
At the algorithmic level, the required number of
Grover queries scales as $N^{1/4}$ for fixed $K$.

To assess physical feasibility, we perform full-sequence
scattering calculations for the cavity-QED implementation,
including finite photon bandwidth and loss.
For suitable system parameters, the prepared states exhibit
appreciable spin squeezing and high target-state fidelity.
Heralding on detection of the reflected photons removes
loss events from accepted runs and improves conditional-state
fidelity, although finite-bandwidth phase errors and imperfect
Dicke selectivity remain.
Implementing the required non-$\pi$ Dicke-selective phases
also demands accurate reflection-phase control and can
impose stricter requirements on resolving neighboring
Dicke resonances.

The remainder of this paper is organized as follows. Section~\ref{sec:central-window} analyzes the spin squeezing and asymptotic scaling of the directly projected $K$-component states, and identifies the smallest parity-allowed targets. Section~\ref{sec:grover-preparation} presents the phase-matched Grover preparation protocols and their query complexity, and extends the construction to the amplitude-reshaped even-$N$ three-component state at unchanged mean-spin contrast. Section~\ref{sec:cavity-qed} describes the cavity-QED implementation and its physical operation counts, and evaluates the complete preparation sequence with finite photon bandwidth and loss. Section~\ref{sec:conclusion} summarizes the main results and discusses prospects for extending the preparation scheme to other physical platforms.

\section{Spin squeezing in symmetric superpositions of neighboring Dicke states}
\label{sec:central-window}

We first define the directly projected targets and obtain their spin moments and Wineland squeezing, and then compare the asymptotic regimes to identify the smallest nontrivial targets allowed by parity.

We work in the permutation-symmetric subspace with total spin $J=N/2$ and Dicke basis $\ket{J,m}$.
The $x$-polarized coherent spin state is
$\ket{\CSS_x}=\sum_{m=-J}^{J}\sqrt{w_m}\ket{J,m}$,
where $w_m=2^{-N}\binom{N}{J+m}$.
Let $\Pi_K$ project onto the $K$ neighboring Dicke components
closest to $m=0$.
Projecting onto this subspace and normalizing gives the target
\begin{equation}
  \ket{\psi_K}
  \equiv
  \frac{\Pi_K\ket{\CSS_x}}{\sqrt{p_K}}.
  \label{eq:intro-target}
\end{equation}
Here $p_K=\sum_{m=-M}^{M}w_m$ is the probability weight of the
selected subspace, with $M=(K-1)/2$.
The allowed Dicke projections require $N+K$ to be odd: $K$ is odd for even $N$ and even for odd $N$.

Reality and reflection symmetry place the mean spin along $x$. Recurrences between neighboring coherent-state weights yield the spin moments and identify $z$ as the direction of minimum transverse variance; the calculation is given in Appendix~\ref{app:direct-moments}.
Writing $\eta\equiv\eta_{J,M}\equiv(J-M)w_M/(Jp_K)$,
where $w_M$ is the probability of the boundary component $m=M$,
the mean spin and minimum transverse variance are
$\langle J_x\rangle=J(1-\eta)$ and
$(\Delta J_z)^2=J(1-K\eta)/2$, respectively.
Equation~\eqref{eq:wineland} then gives the closed-form squeezing
\begin{equation}
  \xi_R^2(N,K)=\frac{1-K\eta}{(1-\eta)^2}.
  \label{eq:central-closed-form}
\end{equation}
This result implies $\xi_R^2<1$ for every parity-allowed target with $2\le K<N+1$, as proved in Appendix~\ref{app:direct-moments}. The two excluded boundaries are the even-$N$ single Dicke state at $K=1$, which has zero mean spin, and the full coherent spin state at $K=N+1$, which is unsqueezed.


For fixed $K\ge2$, the retained coherent-state weights become asymptotically equal as $N$ increases, giving
\begin{equation}
  \xi_R^2
  \sim
  \frac{K^2(K+1)}{3(K-1)}
  N^{-1}.
  \label{eq:fixed-k-squeezing}
\end{equation}
The prefactor increases with integer $K\ge2$, so the smallest allowed fixed support gives the strongest asymptotic squeezing within this directly projected family. When $K$ grows with $N$, the squeezing depends on the support
width relative to the coherent-state distribution.
In the intermediate regime $1\ll K\ll\sqrt N$,
$\xi_R^2\sim K^2N^{-1}/3$.
At $K=c\sqrt N$ with fixed $c>0$, the squeezing approaches a
nonzero constant below unity.
For $K\gg\sqrt N$, the retained support captures essentially
the entire coherent-state distribution and $\xi_R^2\to1$.
The asymptotic derivations are given in
Appendix~\ref{app:asymptotic-scaling}.
Fig.~\ref{fig:squeezing-and-query-counts}(a) shows the finite-$N$ dependence on $K$ and $N$. Strong squeezing is concentrated at small $K$, whereas broad supports approach the coherent-state value $\xi_R^2=1$; the $K=\sqrt N$ reference curve identifies the crossover scale discussed above.

\begin{figure*}[!t]
  \centering
  \includegraphics[width=\textwidth]{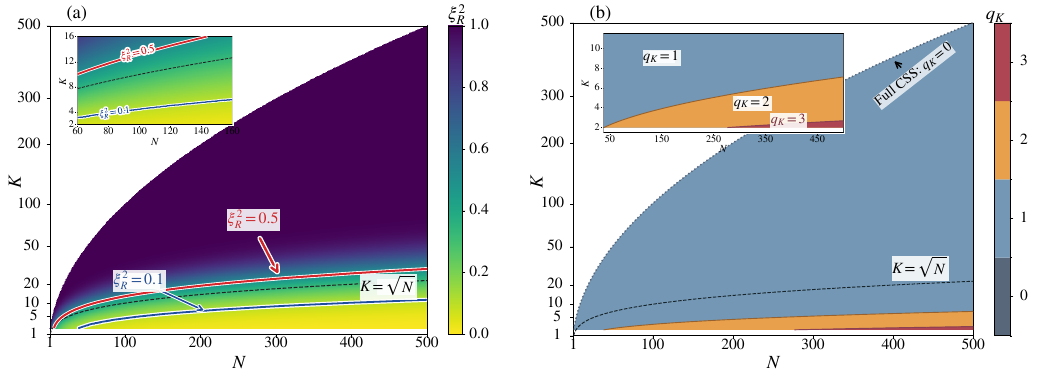}
  \caption{
Ideal squeezing and Grover iteration counts for directly projected
$K$-component targets versus $N$ and $K$.
(a) Wineland parameter $\xi_R^2$, interpolated from exact results
at parity-allowed integer pairs.
(b) Integer iteration count $q_K$ from
Eq.~\eqref{eq:phase-matched-iterations} at the same allowed pairs,
with interpolated boundaries between the integer-count regions.
Both main panels use a square-root $K$ scale;
the insets enlarge the small-$K$ regions on linear axes.
  }
  \label{fig:squeezing-and-query-counts}
\end{figure*}

As shown in Fig.~\ref{fig:squeezing-and-query-counts}(a), the strongest squeezing is achieved for the smallest nontrivial targets allowed by parity. For odd $N$, the smallest allowed target is the equal superposition
$\ket{\psi_2}=(\ket{J,-1/2}+\ket{J,+1/2})/\sqrt2$.
For even $N\ge4$, the directly projected target
$\ket{\psi_{3,\mathrm{dir}}}=\ket{\psi_{K=3}}$
retains $m=-1,0,+1$ with real, positive amplitudes in the ratio
$1:\sqrt{(J+1)/J}:1$.
Equation~\eqref{eq:central-closed-form} gives the corresponding
Wineland parameters for the odd-$N$ and even-$N$ targets,
respectively:
\begin{subequations}
\begin{align}
  &\xi_R^2(N,2)
  =\frac{4N}{(N+1)^2}
  =4N^{-1}+O(N^{-2}),
  \label{eq:odd-minimal-xi}\\
  &\xi_{R,\mathrm{dir}}^2
  =\frac{2(3N+2)}{(N+2)^2}
  =6N^{-1}+O(N^{-2}).
  \label{eq:projected-three-xi}
\end{align}
\end{subequations}

\section{Grover preparation}
\label{sec:grover-preparation}

We now construct phase-matched Grover protocols for the target states identified in Sec.~\ref{sec:central-window}. We first consider the directly projected $K$-component targets and their query complexity, and then examine amplitude redistribution at fixed mean-spin contrast for the even-$N$ three-component state together with its two-stage Grover preparation.

\subsection{Preparation of directly projected targets}

The target in Eq.~\eqref{eq:intro-target} already has the relative Dicke amplitudes of the coherent spin state within the marked subspace. Its preparation therefore requires amplifying the weight of this subspace while preserving those relative amplitudes. We use the phase operations
\begin{subequations}
\label{eq:aa-phase-operations}
\begin{align}
  \chi_{\Pi_K}(\phi)
  &\equiv\mathbb I-(1-e^{\ii\phi})\Pi_K,
  \label{eq:marked-subspace-phase-operation}\\
  \chi_{\CSS}(\phi)
  &\equiv\mathbb I-(1-e^{\ii\phi})\proj{\CSS_x}.
  \label{eq:css-phase-operation}
\end{align}
\end{subequations}
The first operation applies a common phase to the selected Dicke components, while the second applies a phase to the initial-state direction. Their product transfers amplitude toward the target within the two-dimensional target--rest subspace, as detailed in Appendix~\ref{app:phase-matching}.

For $\phi=\pi$, this evolution reduces to the standard Grover rotation illustrated in Fig.~\ref{fig:preparation-overview}(a). An integer number of such rotations generally does not reach the target exactly. Within the arbitrary-phase framework of H{\o}yer~\cite{hoyer2000}, Long's equal-phase construction~\cite{long2001} removes this mismatch by choosing the iteration count and phase as
\begin{subequations}
\label{eq:phase-matched-parameters}
\begin{align}
  q_K
  &=
  \left\lceil
    \frac{\pi}{4\arcsin\sqrt{p_K}}
    -
    \frac{1}{2}
  \right\rceil,
  \label{eq:phase-matched-iterations}\\
  \delta_K
  &=
  2\arcsin
  \left[
    \frac{\sin\!\left(\pi/(4q_K+2)\right)}{\sqrt{p_K}}
  \right].
  \label{eq:phase-matched-phase}
\end{align}
\end{subequations}These parameters make the rest amplitude vanish after $q_K$ steps, giving exact preparation in the ideal phase-oracle model:
\begin{equation}
  \left[
    \chi_{\CSS}(\delta_K)
    \chi_{\Pi_K}(\delta_K)
  \right]^{q_K}
  \ket{\CSS_x}
  =
  e^{\ii\zeta_K}
  \ket{\psi_K},
  \label{eq:exact-aa-central}
\end{equation}
where $e^{\ii\zeta_K}$ is a global phase factor.

\subsection{Query complexity}

For $KN^{-1/2}\to0$, the initial marked-subspace weight decreases with $N$ and increases with $K$. Combining its asymptotic form with Eq.~\eqref{eq:phase-matched-iterations} gives
\begin{equation}
  q_K
  \simeq
  \frac{\pi}{4}
  \left(
    \frac{\pi N}{2}
  \right)^{1/4}
  \frac{1}{\sqrt K}.
  \label{eq:query-complexity}
\end{equation}
Each iteration uses one marked-subspace operation $\chi_{\Pi_K}$, so $q_K$ is also the query count. The scaling follows because amplitude amplification depends on the inverse square root of the initial marked-subspace probability; the intermediate steps are given in Appendix~\ref{app:query-scaling}.

Increasing $K$ thus reduces the query count at the expense of weaker squeezing. In the intermediate regime $1\ll K\ll\sqrt N$, the trade-off is
$  \xi_R^2\sim\pi^5/({1536\,q_K^4)}.$ For fixed $K$, the query count grows as $N^{1/4}$ while the target squeezing improves as $N^{-1}$. At $K=c\sqrt N$ with fixed $c>0$, a bounded query count instead gives a nonzero limiting $\xi_R^2<1$. Figure~\ref{fig:squeezing-and-query-counts}(b) shows the exact integer counts across the $(N,K)$ plane.

\subsection{Amplitude redistribution at fixed mean-spin contrast}
\label{sec:three-dicke}

The mean spin sets the interferometric response, while transverse spin fluctuations determine the noise~\cite{wineland1994}. To isolate squeezing gains from reduced fluctuations, we compare states at fixed mean-spin contrast, $\mathcal C\equiv\lvert\langle\bm J\rangle\rvert/J$. At fixed $N$, this leaves the denominator of the Wineland parameter unchanged. The fixed-mean-spin optimization of S{\o}rensen and M{\o}lmer and the variational formulation of Rojo provide the relevant benchmarks~\cite{sorensen2001,rojo2003}. The odd-$N$ two-component target already minimizes $\xi_R^2$ at its contrast within the full symmetric $J=N/2$ subspace (Appendix~\ref{app:odd-fixed-contrast}). For even $N$, however, the squeezing can be improved by redistributing amplitudes over the same three Dicke components.

The squeezing of Dicke-state superpositions depends on their coefficients and relative phases~\cite{huhu2014}. For even $N\ge4$, we restrict the comparison to real, reflection-symmetric superpositions over $m=-1,0,+1$. Matching the contrast of direct projection admits a second amplitude distribution with greater weight on the central component, as derived in Appendix~\ref{app:three-dicke-details}. This defines the reshaped target
\begin{align}
  \ket{\psi_{3,\mathrm{re}}}
  &\equiv\sqrt{\frac{J+1}{2(3J+1)}}
    \left(\ket{J,-1}+\ket{J,+1}\right)\notag\\
  &\quad+\sqrt{\frac{2J}{3J+1}}\ket{J,0}.
  \label{eq:reshaped-three-state}
\end{align}
It has the same mean spin as the directly projected target but a smaller variance along $z$, giving
\begin{equation}
  \xi_{R,\mathrm{re}}^2
  =\frac{3N+2}{N+2}N^{-1}
  =3N^{-1}+O(N^{-2}).
  \label{eq:reshaped-xi}
\end{equation}
The leading squeezing coefficient is therefore reduced from $6$ to $3$ without changing the Dicke support or the mean-spin response.
For larger odd $K$, redistributing amplitude between the central $m=0$ component and the remaining components, while preserving
the amplitude ratios among the latter, also reduces the Wineland
parameter at fixed mean-spin contrast. The relative improvement
decreases for $1\ll K\ll\sqrt{N}$.

\subsection{Preparation of the reshaped target}
\label{sec:amplitude-reshaping}

The reshaped target requires a larger central-to-side amplitude ratio than the initial coherent spin state. We therefore first modify this ratio and then amplify the resulting three-component target. The reshaping operation is
\begin{equation}
  U_{\mathrm{re}}
  =\chi_{\CSS}(\beta)\chi_0(\alpha),
  \label{eq:reshaping-operation}
\end{equation}
where $\chi_0(\alpha)\equiv\mathbb I-(1-e^{\ii\alpha})\proj{J,0}$ applies a phase to the central Dicke component. The subsequent CSS phase operation changes its amplitude relative to the remaining components while preserving the relative amplitudes among those components. Suitable phases exist for every even $N\ge4$; their calculation is given in Appendix~\ref{app:reshaping-phases}.

\begin{figure}[!t]
  \centering
  \includegraphics[width=\linewidth]{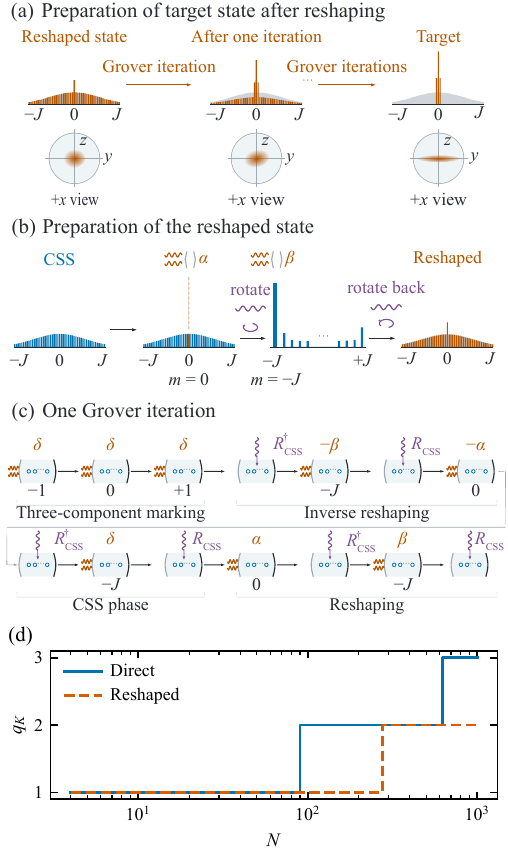}
  \caption{
Reshaped three-component preparation and its physical
operation sequence.
(a) Grover amplification from the reshaped initial state
to the target.
Bars show schematic Dicke-amplitude magnitudes;
the $+x$ views illustrate transverse spin fluctuations.
(b) Initial-state reshaping $U_{\mathrm{re}}$, with single-photon
reflections applying phases $\alpha$ and $\beta$ to the $m=0$ and
$m=-J$ components, respectively; ``rotate'' and ``rotate back''
denote $R_{\mathrm{CSS}}^\dagger$ and $R_{\mathrm{CSS}}$.
(c) One Grover iteration, comprising three-component marking,
inverse reshaping $U_{\mathrm{re}}^\dagger$, a CSS phase operation,
and reshaping $U_{\mathrm{re}}$, with
$\delta=\delta_{\mathrm{re}}$. (d) Exact integer Grover iteration counts for the Direct
and Reshaped three-component protocols at even $N\ge4$.
  }
  \label{fig:reshaped-protocol}
\end{figure}

The resulting initial state $\ket{\psi_{\mathrm{i}}^{(\mathrm{re})}}=U_{\mathrm{re}}\ket{\CSS_x}$ has the required relative target amplitudes within the marked subspace.
Amplitude amplification now uses the phase operation
$\chi_{\mathrm{i}}^{(\mathrm{re})}(\delta)
=U_{\mathrm{re}}\chi_{\CSS}(\delta)U_{\mathrm{re}}^\dagger$
about this reshaped initial state.
Let $p_3^{(\mathrm{re})}$ denote the probability of this initial
state in the three-component marked subspace.
Choosing $q_{\mathrm{re}}$ and $\delta_{\text{re}}$ from
Eqs.~\eqref{eq:phase-matched-iterations}
and~\eqref{eq:phase-matched-phase},
with $p_K$ replaced by $p_3^{(\mathrm{re})}$,
gives exact preparation,
$[\chi_{\mathrm{i}}^{(\mathrm{re})}(\delta_{\text{re}})
\chi_{\Pi_3}(\delta_{\text{re}})]^{q_{\mathrm{re}}}
\ket{\psi_{\mathrm{i}}^{(\mathrm{re})}}
=e^{\ii\zeta'}\ket{\psi_{3,\mathrm{re}}}$.
The corresponding state evolution is illustrated in Fig.~\ref{fig:reshaped-protocol}(a). Initial-state reshaping asymptotically doubles the marked-subspace probability and reduces the required query count by a factor of $\sqrt2$, giving $q_{\mathrm{re}}\simeq0.36\,N^{1/4}$. Figure~\ref{fig:reshaped-protocol}(d) compares the exact integer counts and shows the delayed step-up transitions of the reshaped protocol.

\section{Cavity-QED implementation and performance}
\label{sec:cavity-qed}

We now consider a cavity quantum electrodynamics (cavity QED) implementation, in which an atomic ensemble is dispersively coupled to a cavity mode~\cite{li2022,reiserer2015}. The cavity response depends on the atomic excitation number, allowing reflected photons to imprint the selective phases required by the Grover protocols~\cite{schleier2010,chen2015,nagib2025prl,reiserer2014}. We describe how photon reflections and global rotations realize these protocols, count the physical operations, and assess the complete sequence in the presence of finite photon bandwidth and loss. The performance comparison focuses on the directly projected and reshaped three-component targets for even $N\ge4$.

\subsection{Dicke-selective phase operations}
\label{sec:cavity-phase-operations}

In the dispersive regime, the effective atom--cavity interaction is described by
\begin{equation}
    H=\hbar\Omega\hat n\hat n_c,
    \label{eq:dispersive-hamiltonian}
\end{equation}
where $\hat n=J+J_z$ is the atomic excitation-number operator and $\hat n_c$ is the cavity photon-number operator. The shift per atomic excitation is $\Omega=g_c^2/\Delta_a$, with single-atom coupling $g_c$ and detuning $\Delta_a=\omega_0-\omega_a$ between the bare-cavity resonance and the atomic transition. For a Dicke state with excitation number $n=J+m$, the interaction shifts the cavity resonance to $\omega_0+n\Omega$. Thus, the phase acquired upon photon reflection depends on the atomic Dicke component. We denote the corresponding excitation-number state by $\ket n\equiv\ket{J,m=n-J}$.

In the ideal one-sided lossless limit, a photon centered at
$\bar\omega_p=\omega_0+n\Omega+\kappa\cot(\phi/2)$
imprints phase $\phi$ on the target component, where $\kappa$
is the cavity field-amplitude decay rate. Resonant reflection gives a $\pi$ phase, while non-$\pi$ operations require an off-resonant photon frequency. The reflection amplitudes and the frequency-selection condition are given in Appendix~\ref{app:cavity-response}.
The photon must also leave the other Dicke components nearly unchanged. This selectivity is controlled by the spectral resolution $d=\Omega/\kappa$: large $|d|$ improves the discrimination of neighboring Dicke resonances. For non-$\pi$ phases, the operating frequency can lie closer to a neighboring resonance, making selectivity more demanding. Narrow photon bandwidth and low scattering loss are also required to approach the ideal operation.

In the ideal selective limit, the photon implements
$\chi_m(\phi)=\mathbb I-(1-e^{\ii\phi})\proj{J,m}$, with $m=n-J$.
Applying these phases to the selected Dicke components realizes the marked-subspace operation, while global rotations convert a phase on $\ket{n=0}$ into a CSS phase operation:
\begin{subequations}
\label{eq:cavity-operator-map}
\begin{align}
  \chi_{\Pi_K}(\phi)
  &=\prod_{m=-M}^{M}\chi_m(\phi),
  \label{eq:cavity-map-k}\\
  \chi_{\CSS}(\phi)
  &=R_{\CSS}\chi_{-J}(\phi)R_{\CSS}^\dagger,
  \label{eq:cavity-map-css}
\end{align}
\end{subequations}
where the global rotation $R_{\CSS}\equiv R_y(-\pi/2)$, with $R_y(\vartheta)\equiv e^{-\ii\vartheta J_y}$, maps $\ket{n=0}$ to $\ket{\CSS_x}$. Figure~\ref{fig:reshaped-protocol}(b) shows the initial reshaping sequence. Combining this sequence with its inverse and the CSS phase operation realizes the conjugation defining $\chi_{\mathrm{i}}^{(\mathrm{re})}$ in Sec.~\ref{sec:amplitude-reshaping}, giving the complete iteration in Fig.~\ref{fig:reshaped-protocol}(c).

\subsection{Physical operation counts}
\label{sec:Physical operation counts}
\label{sec:physical-resources}

\begin{figure}[!htbp]
  \centering
  \includegraphics[width=\linewidth]{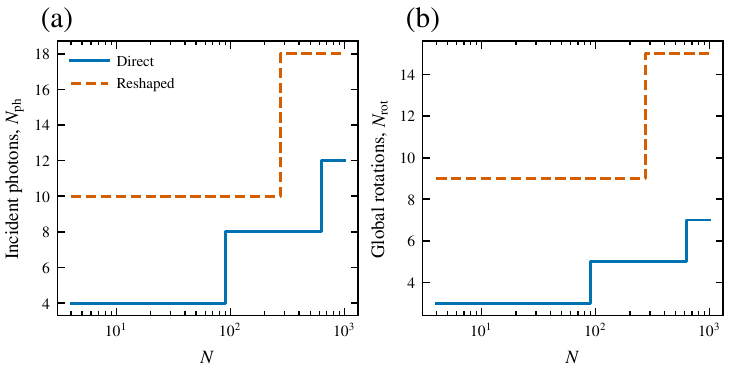}
  \caption{
Physical resources for the cavity-QED implementation
of the Direct and Reshaped three-component protocols:
(a) incident photons and (b) global rotations.
Counts include initial preparation and use the exact
integer Grover counts.
  }
  \label{fig:operation-counts}
\end{figure}

Each $\chi_m$ uses one incident photon, and each $R_{\CSS}$ or $R_{\CSS}^\dagger$ counts as one global rotation. Table~\ref{tab:operation-counts} summarizes the elementary and composite operations.
With initial-state preparation included and adjacent inverse
rotations canceled, the even-$N$ direct-projection protocol
uses $4q_{\mathrm{dir}}$ photons and $1+2q_{\mathrm{dir}}$
global rotations, whereas the reshaped protocol uses
$2+8q_{\mathrm{re}}$ photons and $3+6q_{\mathrm{re}}$
global rotations.
The odd-$N$ two-component protocol requires $N_{\mathrm{ph}}^{(2)}=3q_2$ photons and $N_{\mathrm{rot}}^{(2)}=1+2q_2$ rotations. All three protocols retain the $N^{1/4}$ resource scaling. Reshaping can reduce the Grover iteration count, but each of its iterations uses more photons and global rotations than a direct-projection iteration. Figure~\ref{fig:operation-counts} compares the exact total counts for the two three-component protocols; these counts set the sequence lengths used in the scattering analysis below.

\begin{table}[b]
    \caption{Ideal physical operation counts for the cavity-QED building blocks.}
    \label{tab:operation-counts}
    \begin{ruledtabular}
        \begin{tabular}{lcc}
            Operation & Photons & Global rotations \\
            \hline
            $\chi_m(\phi)$
            & 1 & 0 \\
            $\chi_{\mathrm{CSS}}(\phi)$
            & 1 & 2 \\
            $\chi_{\Pi_2}(\phi)$
            & 2 & 0 \\
            $\chi_{\Pi_3}(\phi)$
            & 3 & 0 \\
            $U_{\mathrm{re}}$ or $U_{\mathrm{re}}^\dagger$
            & 2 & 2 \\
            $\chi_{\mathrm{i}}^{(\mathrm{re})}(\phi)$
            & 5 & 6 \\
        \end{tabular}
    \end{ruledtabular}
\end{table}

\subsection{Performance with finite photon bandwidth and loss}
\label{sec:realistic-performance}

We assess the physical preparation by propagating the complete sequence of photon reflections and global rotations within the symmetric $J=N/2$ subspace. Each reflection is described by a scattering map diagonal in excitation number and averaged over the photon spectrum. The unheralded state includes all outgoing channels, whereas the heralded state is conditioned on reflection of every photon in the sequence. Heralding excludes loss events, but phase errors from finite bandwidth and imperfect Dicke selectivity remain.

For each normalized final state $\rho_{\mathrm{fin}}$, we evaluate the target-state fidelity $F=\bra{\psi_{\mathrm t}}\rho_{\mathrm{fin}}\ket{\psi_{\mathrm t}}$ and the Wineland parameter in Eq.~\eqref{eq:wineland}. The latter is computed from the mean-spin vector and the minimum transverse variance, without assuming that the mean spin remains along $x$ or that the squeezing direction remains along $z$. The scattering maps, sequence propagation, and evaluation of this variance are detailed in Appendix~\ref{app:finite-bandwidth}.

The numerical calculations use Gaussian photon wave packets with root-mean-square spectral width $\sigma$ and relative bandwidth $w=\sigma/\kappa=0.1$. We take $\kappa/\gamma=1$ and a one-sided cavity with atomic spontaneous emission retained, where $\gamma$ is the atomic polarization decay rate. The atom--cavity coupling relative to decay is characterized by the single-atom cooperativity $C=g_c^2/(\kappa\gamma)$. At each $N$ and $C$, a common $d$ for the complete sequence is optimized separately for each protocol and heralding condition to maximize $F$. The squeezing is evaluated at the same operating point.

\begin{figure}[!t]
  \centering
  \includegraphics[width=\linewidth]{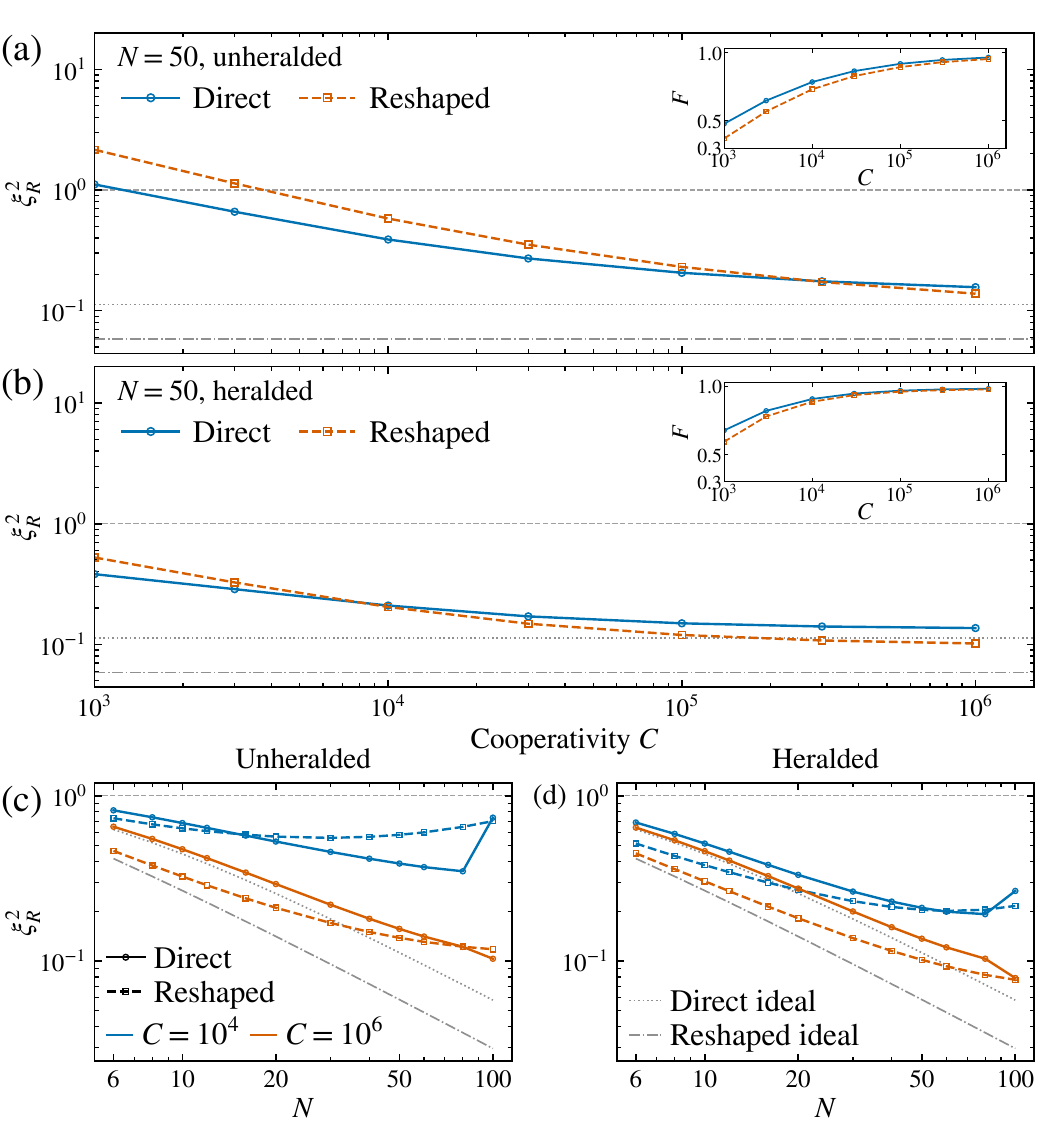}
  \caption{
Full-sequence Wineland squeezing with finite photon bandwidth
and loss at $w=0.1$ and $\kappa/\gamma=1$.
(a),(b) Cooperativity scans for $N=50$, with the corresponding
target-state fidelities shown in the insets.
(c),(d) Particle-number scans at fixed cooperativity.
Protocol line styles are shared across all panels.
At each point, $d$ is optimized to maximize $F$,
and $\xi_R^2$ is evaluated at the same operating point.
Gray dashed lines mark $\xi_R^2=1$.
  }
  \label{fig:realistic-performance}
\end{figure}

Figure~\ref{fig:realistic-performance} shows that the complete sequence retains $\xi_R^2<1$ for suitable cavity parameters. For $N=50$, panels (a) and (b) show that increasing $C$ improves both target-state fidelity and squeezing. The reshaped protocol recovers the squeezing advantage of its ideal target at lower cooperativity with heralding than without it.

Figures~\ref{fig:realistic-performance}(c) and \ref{fig:realistic-performance}(d) show the particle-number dependence at fixed cooperativity. The direct-projection curves turn upward at $N=100$ for $C=10^4$, consistent with the increase from one to two Grover iterations near $N\approx90$ in Fig.~\ref{fig:reshaped-protocol}(d) and the corresponding increase from four to eight photons in Fig.~\ref{fig:operation-counts}(a). In this regime, the error cost of a longer physical sequence can outweigh the improvement in ideal-target squeezing with increasing $N$. The larger initial overlap of the reshaped protocol postpones its first step-up transition to $N\approx270$.

Table~\ref{tab:full-sequence-benchmarks} summarizes representative full-sequence results, including the heralding success probability $P_{\mathrm{succ}}$ and the spectral resolution $d_{\mathrm{opt}}$ that maximizes $F$. Across these benchmarks, direct projection gives the higher target-state fidelity, whereas the ordering of $\xi_R^2$ depends on $N$, $C$, and heralding. At $N=50$ and $C=10^4$, the reshaped protocol gives weaker squeezing in the unheralded output but comparable squeezing after heralding. At $N=50$ and $C=10^6$, it gives stronger squeezing in both branches. For $N=50$, direct projection also gives a higher heralding success probability and a lower mean incident-photon consumption per successful preparation over the range studied in Appendix~\ref{app:heralding-cost}.

\begin{table}[t]
\caption{Representative full-sequence benchmarks at $w=0.1$
and optimized $d$. D and R denote the direct-projection and
reshaped protocols, respectively. $P_{\mathrm{succ}}$ is
reported only for heralded outputs.}
\label{tab:full-sequence-benchmarks}
\centering

\setlength{\tabcolsep}{4pt}
\renewcommand{\arraystretch}{1.10}

\begin{tabular}{cccccc}
    \hline\hline
    Heralding & Protocol & $F$ & $\xi_R^2$
    & $P_{\mathrm{succ}}$ & $d_{\mathrm{opt}}$ \\
    \hline

    \multicolumn{6}{c}{$N=10,\quad C=10^{4}$} \\[2pt]
    \multirow{2}{*}{Unheralded}
      & D & 0.889 & 0.683 & ---   & 12.913 \\
      & R & 0.839 & 0.635 & ---   & 16.170 \\
    \multirow{2}{*}{Heralded}
      & D & 0.973 & 0.514 & 0.803 & 27.756 \\
      & R & 0.953 & 0.380 & 0.708 & 35.519 \\[4pt]

    \multicolumn{6}{c}{$N=50,\quad C=10^{4}$} \\[2pt]
    \multirow{2}{*}{Unheralded}
      & D & 0.786 & 0.389 & ---   & 8.599  \\
      & R & 0.732 & 0.580 & ---   & 8.945  \\
    \multirow{2}{*}{Heralded}
      & D & 0.909 & 0.210 & 0.814 & 13.594 \\
      & R & 0.889 & 0.204 & 0.706 & 15.325 \\[4pt]

    \multicolumn{6}{c}{$N=50,\quad C=10^{6}$} \\[2pt]
    \multirow{2}{*}{Unheralded}
      & D & 0.966 & 0.157 & ---   & 23.022 \\
      & R & 0.955 & 0.138 & ---   & 24.728 \\
    \multirow{2}{*}{Heralded}
      & D & 0.984 & 0.137 & 0.952 & 48.050 \\
      & R & 0.979 & 0.102 & 0.918 & 54.784 \\

    \hline\hline
\end{tabular}
\end{table}

\section{Conclusion and Outlook}
\label{sec:conclusion}

We have shown that phase-matched Grover amplitude amplification can exactly prepare spin-squeezed superpositions of a few neighboring Dicke states symmetrically distributed about $m=0$ in the ideal phase-oracle model. For fixed $K$, the directly projected targets exhibit $N^{-1}$ Wineland squeezing, with the smallest nontrivial targets allowed by parity giving the strongest asymptotic squeezing within this family. The odd-$N$ two-component target minimizes $\xi_R^2$ at its mean-spin contrast within the full symmetric $J=N/2$ subspace. For even $N$, redistributing amplitudes among the same three Dicke components halves the leading coefficient of $\xi_R^2$ while preserving the mean-spin contrast. For fixed $K$, the Grover query count scales as $N^{1/4}$. Full-sequence scattering calculations for a cavity-QED implementation using global rotations and Dicke-selective phase operations show that appreciable spin squeezing and high target-state fidelity can be achieved for suitable system parameters despite finite photon bandwidth and loss. Heralding on photon reflection improves the conditional-state fidelity, while finite-bandwidth phase errors and imperfect Dicke selectivity remain the main sources of error.

The implementation considered here uses reflected single photons in an optical cavity to provide the Dicke-selective phase shifts required for Grover amplitude amplification, making the preparation sensitive to finite photon bandwidth and losses accumulated over repeated scattering. In Rydberg-atom ensembles, these phase operations could instead be mediated by an auxiliary atom, avoiding the need for repeated optical photon reflections~\cite{anikeeva2021}. Superconducting circuit-QED systems offer another route through collective dispersive coupling and microwave control of Dicke-state energy shifts~\cite{boissonneault2009,zhang2024,blais2021}. Whether these alternatives introduce other significant errors remains to be investigated.

Further opportunities arise from the reflection symmetry of our target states, which are superpositions of a few neighboring Dicke states near $m=0$. Interactions producing $m^2$-dependent energy shifts could allow components with opposite spin projections to be addressed together, potentially reducing the number of distinct frequencies required for Dicke-selective phase operations. Such interactions occur in two-component Bose--Einstein condensates, while controllable Dicke-state energy shifts have also been explored in cavity-coupled systems~\cite{jin2008,sorensen2001bec,riedel2010,zhang2024,wang2026dicke}. Combined with global rotations, this symmetry-based control may also simplify the CSS reflection within the reflection-symmetric subspace for odd $N$.. Its compatibility with the complete phase-matched Grover protocol remains to be established.

\begin{acknowledgments}
This work is supported by the National Natural Science Foundation of China
(Grant Nos.~12574082 and E31Q02BG),
the Chinese Academy of Sciences
(Grant Nos.~E0SEBB11 and E27RBB11),
the Quantum Science and Technology---National Science and Technology Major Project
(Grant No.~2021ZD0302300),
and the Chinese Academy of Sciences Project for Young Scientists in Basic Research
(Grant No.~YSBR-090).
\end{acknowledgments}

\section*{Data Availability}

There are no publicly available research data or software supporting this manuscript. Requests for further information or data should be sent to the authors.

\appendix

\section{Analytic properties of the spin-squeezed states}
\label{app:analytic-properties}

\subsection{Spin moments of the $K$-component states}
\label{app:direct-moments}

The collective spin operators are $J_\alpha\equiv\frac12\sum_{j=1}^{N}\sigma_\alpha^{(j)}$ for $\alpha\in\{x,y,z\}$, and the Dicke basis satisfies $J_z\ket{J,m}=m\ket{J,m}$. The $x$-polarized coherent spin state has the expansion $\ket{\CSS_x}=\sum_{m=-J}^{J}c_m\ket{J,m}$, with
\begin{equation}
  c_m=2^{-J}\sqrt{\binom{2J}{J+m}}.
  \label{eq:css-dicke-amplitudes-app}
\end{equation}
The projector in Eq.~\eqref{eq:intro-target} is $\Pi_K=\sum_{m=-M}^{M}\proj{J,m}$. With $w_m=c_m^2$, its coherent-state weight is $p_K=\sum_{m=-M}^{M}w_m$, and the target amplitudes are $c_m/\sqrt{p_K}$ on the retained components. The coherent-state amplitudes satisfy the adjacent-amplitude recurrence
$
  \frac{c_{m+1}}{c_m}=\sqrt{\frac{J-m}{J+m+1}}.
$
The standard ladder-operator matrix elements then give

\begin{subequations}
\label{eq:coherence-relations}
\begin{align}
  c_m c_{m+1}\langle J,m+1|J_+|J,m\rangle
  &= (J-m)w_m,
  \label{eq:one-step-coherence}\\
  c_m c_{m+2}\langle J,m+2|J_+^2|J,m\rangle
  &= (J-m-1)\notag\\
  &\quad\times (J-m)w_m.
  \label{eq:two-step-coherence}
\end{align}
\end{subequations}

Using $J_x=(J_++J_-)/2$ and
Eq.~\eqref{eq:one-step-coherence}, the first moment becomes
\begin{equation}
p_K\langle J_x\rangle
=
\sum_{m=-M}^{M-1}(J-m)w_m
=
Jp_K-(J-M)w_M.
\label{eq:jx-boundary}
\end{equation}
The last equality follows from the symmetry
$\sum_{m=-M}^{M}mw_m=0$.
Using the boundary quantity $\eta$ defined in Sec.~\ref{sec:central-window} gives the mean spin stated there.

The covariance matrix is defined as
\begin{equation}
  \Gamma_{\alpha\beta}
  \equiv
  \frac{1}{2}
  \langle J_\alpha J_\beta+J_\beta J_\alpha\rangle
  -
  \langle J_\alpha\rangle\langle J_\beta\rangle.
  \label{eq:covariance-definition}
\end{equation}
For the $z$-component second moment, define
$q_m=-m^2/2+(1/4-J/2)m+J/4$.
The weight recurrence gives
\begin{equation}
  \left(m^2-\frac J2\right)w_m
  =
  q_{m+1}w_{m+1}-q_mw_m.
\end{equation}
Summing over the selected Dicke components and using
$w_{-M}=w_M$ and
$w_{M+1}=(J-M)w_M/(J+M+1)$ yields
\begin{equation}
  \sum_{m=-M}^{M}m^2w_m
  =
  \frac J2p_K
  -
  \left(M+\frac12\right)(J-M)w_M.
\end{equation}
Since $K=2M+1$ and $\langle J_z\rangle=0$, division by $p_K$
gives $\Gamma_{zz}=\frac{J}{2}(1-K\eta)$, as stated in Sec.~\ref{sec:central-window}.

The identities
$J_x^2+J_y^2=J(J+1)-J_z^2$ and
$J_x^2-J_y^2=(J_+^2+J_-^2)/2$
reduce the calculation of the $x$- and $y$-component second moments to Eq.~\eqref{eq:two-step-coherence}.
Completing the resulting symmetric finite sum gives
\begin{equation}
  \langle J_x^2-J_y^2\rangle
  =
  J\left(J-\frac12\right)
  -
  J\left(2J+M-\frac12\right)\eta.
\end{equation}
Combining this result with the angular-momentum identities and
subtracting $\langle J_x\rangle^2$ gives the remaining diagonal variances
\begin{subequations}
\label{eq:central-moments}
\begin{align}
  \Gamma_{yy}
  &=
  \frac{J}{2}
  \left[
    1+2(J+M)\eta
  \right],
  \label{eq:central-gammayy}\\
  \Gamma_{xx}
  &=
  \frac{J\eta}{2}
  \left(
    2J+1-2J\eta
  \right).
  \label{eq:central-gammaxx}
\end{align}
\end{subequations}
Reality and reflection symmetry give $\langle J_y\rangle=\langle J_z\rangle=0$ and eliminate the off-diagonal covariances. Comparing the transverse variances yields
\begin{equation}
  \Gamma_{yy}-\Gamma_{zz}
  =\frac{J\eta}{2}(2J+4M+1)>0
  \label{eq:minimum-variance-axis-app}
\end{equation}
for $2\le K<N+1$, so the minimum transverse variance is $\Gamma_{zz}=(\Delta J_z)^2$.

For every parity-allowed choice with $2\le K<N+1$, the target contains an adjacent pair of Dicke components with positive amplitudes and therefore has $\langle J_x\rangle>0$. Since $\eta>0$ for $M<J$, the relation $\langle J_x\rangle=J(1-\eta)$ then implies $0<\eta<1$, while nonnegativity of the variance gives $1-K\eta\ge0$. The difference between the denominator and numerator in Eq.~\eqref{eq:central-closed-form} is
\begin{equation}
  (1-\eta)^2-(1-K\eta)
  =(K-2)\eta+\eta^2>0,
  \label{eq:squeezing-positivity-app}
\end{equation}
which proves $\xi_R^2<1$ throughout this range.

\subsection{Asymptotic scaling with $K$}
\label{app:asymptotic-scaling}

Using the local central-limit approximation for the symmetric
binomial distribution~\cite{petrov1964}, near the center one has
\begin{equation}
  w_m
  \sim
  \sqrt{\frac{2}{\pi}}N^{-1/2}
  \exp\!\left(-2m^2N^{-1}\right),
  \label{eq:local-central-limit}
\end{equation}
uniformly for $|m|N^{-1/2}\to0$.
This form applies to both fixed $K$ and
$1\ll K\ll\sqrt N$, while the regime $K=c\sqrt N$
is described by the corresponding integrated Gaussian limit.

For $KN^{-1/2}\to0$, the selected weights are equal to leading order, giving
\begin{equation}
  p_K
  \sim
  \sqrt{\frac{2}{\pi}}KN^{-1/2}.
  \label{eq:subcritical-projection-probability}
\end{equation}
For fixed $K$, the finite-sum identity
$K^{-1}\sum_{m=-M}^{M}m^2=(K^2-1)/12$,
together with
$\langle J_x\rangle/J\to(K-1)/K$,
gives the fixed-$K$ squeezing law in Eq.~\eqref{eq:fixed-k-squeezing}.

When $1\ll K\ll\sqrt N$, the retained Dicke-state weights remain
asymptotically equal, $\mathcal C\to1$, and
$(\Delta J_z)^2\sim K^2/12$, giving $\xi_R^2\sim K^2N^{-1}/3$.
For $K=N^\alpha$ with $0<\alpha<1/2$, this corresponds to
$\xi_R^2\sim N^{2\alpha-1}/3$.

In the critical regime $K=c\sqrt N$, with $z=2mN^{-1/2}$, the Gaussian limit gives
\begin{subequations}
\label{eq:critical-gaussian-limits}
\begin{align}
  p_K&\longrightarrow\operatorname{erf}(c/\sqrt2),
  \label{eq:critical-projection-limit}\\
  \mathbb E[z^2\mid |z|\le c]
&\longrightarrow
1-\sqrt{\frac{2}{\pi}}\,
\frac{ce^{-c^2/2}}{\operatorname{erf}(c/\sqrt2)}.
  \label{eq:critical-conditional-moment}
\end{align}
\end{subequations}
Because $4(\Delta J_z)^2N^{-1}$ approaches the limiting conditional variance in Eq.~\eqref{eq:critical-conditional-moment} and $\mathcal C\to1$, one obtains the critical-regime squeezing limit described in Sec.~\ref{sec:central-window}.
For $K\gg\sqrt N$, the omitted tails vanish,
$p_K\to1$, $\eta\to0$, and $\xi_R^2\to1$. Only bounded $K$ gives squeezing bounded above and below by positive multiples of $N^{-1}$ within the
directly projected family.

\subsection{Fixed-contrast optimality of the two-component state}
\label{app:odd-fixed-contrast}

For odd $N$, the total spin $J=N/2$ is half-integer.
For any unit vector $\mathbf n$, the eigenvalues of
$J_{\mathbf n}$ are half-integers, so
$J_{\mathbf n}^2\ge\mathbb I/4$.
For a direction $\mathbf n$ perpendicular to the mean-spin vector,
$\langle J_{\mathbf n}\rangle=0$ and therefore
$(\Delta J_{\mathbf n})^2\ge1/4$.
Hence every state in the symmetric $J=N/2$ subspace with nonzero
mean spin satisfies
$(\Delta J_\perp)^2_{\min}\ge1/4$.

For the two-component state $\ket{\psi_2}$, the equal amplitudes give
\begin{equation}
\label{eq:odd-minimal-moments-app}
  \langle J_x\rangle = \frac{N+1}{4}, \qquad (\Delta J_z)^2 = \frac{1}{4}.
\end{equation}
The mean spin points along $x$, so the state attains the transverse-variance bound. Substitution into Eq.~\eqref{eq:wineland} gives Eq.~\eqref{eq:odd-minimal-xi}.
At fixed mean-spin contrast
$\mathcal C=|\langle\mathbf J\rangle|/J$,
the denominator of the Wineland parameter is fixed.
The state $\ket{\psi_2}$ therefore minimizes
$\xi_R^2$ at its mean-spin contrast within the full symmetric
$J=N/2$ subspace.

\subsection{Amplitude redistribution in the three-component state}
\label{app:three-dicke-details}

For even $N\ge4$ ($J\ge2$), consider the real, reflection-symmetric family with $0<y<1$,
\begin{align}
  \ket{\phi(y)}
  &\equiv\sqrt{\frac{1-y}{2}}
    \left(\ket{J,-1}+\ket{J,+1}\right)\notag\\
  &\quad+\sqrt y\ket{J,0}.
  \label{eq:three-dicke-family}
\end{align}
For this state, only the links
$m=-1\leftrightarrow0\leftrightarrow+1$ contribute to
$\langle J_x\rangle=\sqrt{2J(J+1)y(1-y)}$.
Their matrix elements give the squared mean-spin contrast
\begin{equation}
  \mathcal C^2
  =
  \frac{2(J+1)}{J}
  y(1-y),
  \label{eq:three-dicke-contrast}
\end{equation}
while reflection symmetry gives $\langle J_z\rangle=0$ and
$(\Delta J_z)^2=1-y$.
Using
$J_y^2=[J(J+1)-J_z^2]/2-(J_+^2+J_-^2)/4$,
the only two-step coherence joins $m=-1$ and $m=+1$, giving
\begin{equation}
  \Gamma_{yy}-\Gamma_{zz}
  =
  \frac{J(J+1)(1+y)}{4}
  -
  \frac{3(1-y)}{2}.
\end{equation}
This quantity is positive on the larger-$y$ fixed-contrast branch
for $J>1$; in particular, it is positive at
$y_{\mathrm{re}}=2J/(3J+1)$, so the minimum transverse variance is attained along $z$ and Eq.~\eqref{eq:wineland} reduces to
\begin{equation}
  \xi_R^2(y)
  =
  \frac{1}{(J+1)y}.
  \label{eq:three-dicke-xi}
\end{equation}
The directly projected state has $y_{\mathrm{dir}}=(J+1)/(3J+1)$ and contrast $\mathcal C_{\mathrm{dir}}=2(J+1)/(3J+1)$. Since Eq.~\eqref{eq:three-dicke-contrast} is invariant under $y\leftrightarrow1-y$, fixing $\mathcal C=\mathcal C_{\mathrm{dir}}$ gives the two roots
\begin{equation}
  y_\pm
  =
  \frac{1}{2}
  \left[
    1
    \pm
    \sqrt{
      1
      -
      \frac{2J}{J+1}
      \mathcal C^2
    }
  \right].
  \label{eq:contrast-roots}
\end{equation}
The smaller root is $y_-=y_{\mathrm{dir}}$, while the larger root is $y_+=y_{\mathrm{re}}=1-y_{\mathrm{dir}}=2J/(3J+1)$ and gives the target in Eq.~\eqref{eq:reshaped-three-state}. Both states have $\langle J_x\rangle=2J(J+1)/(3J+1)$, whereas their $z$-axis variances are $(\Delta J_z)^2_{\mathrm{dir}}=2J/(3J+1)$ and $(\Delta J_z)^2_{\mathrm{re}}=(J+1)/(3J+1)$. The smaller variance of the reshaped state therefore gives Eq.~\eqref{eq:reshaped-xi}, with the exact squeezing ratio
\begin{equation}
  \frac{\xi_{R,\mathrm{dir}}^2}{\xi_{R,\mathrm{re}}^2}
  =\frac{2N}{N+2}\longrightarrow2.
  \label{eq:reshaped-squeezing-ratio-app}
\end{equation}

\section{Analysis of the amplification protocol}
\label{app:phase-matching}

\subsection{Phase matching and query scaling}
\label{app:query-scaling}

Let $\ket T$ and $\ket B$ be normalized target and rest
directions, with
$\ket{\psi_i}=\sin\theta\ket T+\cos\theta\ket B$.
Define
$\chi_T(\delta)=\mathbb I-(1-e^{\ii\delta})\proj T$ and
$\chi_i(\delta)=\mathbb I-(1-e^{\ii\delta})\proj{\psi_i}$.
This specialization follows Long's equal-phase construction
\cite{long2001} within the general arbitrary-phase framework
\cite{hoyer2000}.
Setting $\sin\theta=\sqrt{p_K}$ gives
Eqs.~\eqref{eq:phase-matched-iterations} and
\eqref{eq:phase-matched-phase}.
The ceiling in Eq.~\eqref{eq:phase-matched-iterations}
ensures that the argument of the arcsine in
Eq.~\eqref{eq:phase-matched-phase} does not exceed unity,
and the rest component vanishes after the corresponding
$q$ iterations.

For the directly projected targets, the two directions are $\ket T=\ket{\psi_K}$ and $\ket B=\ket{\psi_{K,\perp}}$, where the latter is normalized and orthogonal to the image of $\Pi_K$. The coherent spin state decomposes as
\begin{equation}
  \ket{\CSS_x}
  =\sqrt{p_K}\ket{\psi_K}
   +\sqrt{1-p_K}\ket{\psi_{K,\perp}}.
  \label{eq:target-rest-decomposition-app}
\end{equation}
Thus $\Pi_K\ket{\CSS_x}=\sqrt{p_K}\ket{\psi_K}$, while $\chi_{\Pi_K}(\phi)\ket{\psi_K}=e^{\ii\phi}\ket{\psi_K}$ and $\chi_{\Pi_K}(\phi)\ket{\psi_{K,\perp}}=\ket{\psi_{K,\perp}}$. The marked-subspace phase operation preserves the target's internal relative amplitudes, giving the two-dimensional evolution used in Sec.~\ref{sec:grover-preparation}.

For $KN^{-1/2}\to0$, the projection probability satisfies $p_K\sim K\sqrt{2/\pi}N^{-1/2}\to0$. Expanding $\arcsin\sqrt{p_K}$ in Eq.~\eqref{eq:phase-matched-iterations} gives $q_K\simeq\pi/(4\sqrt{p_K})$ to leading order. Substitution yields Eq.~\eqref{eq:query-complexity}. In particular, $q_{\mathrm{dir}}\simeq0.51\,N^{1/4}$ for $K=3$. In the intermediate regime $1\ll K\ll\sqrt N$, solving Eq.~\eqref{eq:query-complexity} for $K$ gives $K\simeq\pi^{5/2}\sqrt N/(16\sqrt2\,q_K^2)$. Substituting this into $\xi_R^2\sim K^2(3N)^{-1}$
gives $\xi_R^2\sim\pi^5/(1536q_K^4)$. For fixed $K$, the same query scaling prepares targets with $N^{-1}$ squeezing and a finite, positive asymptotic coefficient. For $K=c\sqrt N$, Eq.~\eqref{eq:critical-projection-limit} instead gives a nonzero limiting overlap, so the exact integer count remains bounded.

For $KN^{-1/2}\to0$, increasing $K$ reduces the query count, while the
$K$ Dicke-selective phase operations required in each marked-subspace
operation give a sequential phase-operation count
$Kq_K\sim(\pi/4)(\pi N/2)^{1/4}\sqrt K$.
The corresponding physical operation counts are summarized in
Sec.~\ref{sec:physical-resources}.

\subsection{Initial-state reshaping}
\label{app:reshaping-phases}

The coherent-state amplitudes satisfy $c_0/c_1=\sqrt{(J+1)/J}$, while the reshaped target requires a central-to-side amplitude ratio of $2\sqrt{J/(J+1)}$. The required enhancement is therefore
\begin{equation}
  g_{\mathrm{amp}}
  \equiv\frac{\langle J,0|\psi_{3,\mathrm{re}}\rangle}
                {\langle J,1|\psi_{3,\mathrm{re}}\rangle}
        \frac{c_1}{c_0}
  =\frac{2N}{N+2}.
  \label{eq:reshaping-enhancement-app}
\end{equation}
Decompose the coherent spin state as $\ket{\CSS_x}=s_0\ket{J,0}+c_\perp\ket r$, with $s_0=\sqrt{w_0}$ and $c_\perp=\sqrt{1-w_0}$. Here $\ket r$ is normalized, orthogonal to $\ket{J,0}$, and retains the relative coherent-state amplitudes among all $m\ne0$ components. To determine the phases in Eq.~\eqref{eq:reshaping-operation}, write $s=s_0$, $c=c_\perp$, $g=g_{\mathrm{amp}}$, $A=e^{\ii\alpha}$, and $B=e^{\ii\beta}$.
Relative to the input amplitudes $s$ and $c$, the two components acquire
\begin{subequations}
\begin{align}
  F_0
  &=
  c^2(A-1)+B(c^2+s^2A),\\
  F_\perp
  &=
  s^2(1-A)+B(c^2+s^2A).
\end{align}
\end{subequations}
The condition $F_0=gF_\perp$ gives
\begin{equation}
  B=\frac{(c^2+gs^2)(A-1)}{(g-1)(c^2+s^2A)}.
  \label{eq:beta-from-alpha-app}
\end{equation}
Imposing $|B|=1$ and choosing $0<\alpha<\pi$ then yields
\begin{equation}
  \alpha=2\arcsin\!\left[
    \frac{g-1}{2\sqrt{c^2+g^2s^2}}
  \right].
  \label{eq:alpha-app}
\end{equation}
The identity
$(c^2+gs^2)^2+(g-1)^2c^2s^2=c^2+g^2s^2$
guarantees that Eq.~\eqref{eq:beta-from-alpha-app} has unit modulus,
and $\beta=\arg B\pmod{2\pi}$.
For even $N\ge4$, $1<g<2$ and $0<s^2<1$, so both phases exist.
For large $N$, the phases approach $\alpha\to\pi/3$ and $\beta\to2\pi/3$.

The amplitudes transform as $s_0\mapsto s_0F_0$ and $c_\perp\mapsto c_\perp F_\perp$. Since $F_0/F_\perp=g_{\mathrm{amp}}$, the normalized output is
\begin{equation}
    \begin{aligned}
  \ket{\psi_{\mathrm{i}}^{(\mathrm{re})}}
  &\equiv U_{\mathrm{re}}\ket{\CSS_x}\\
  &=e^{\ii\zeta_{\mathrm{re}}}
  \frac{g_{\mathrm{amp}}s_0\ket{J,0}+c_\perp\ket r}
       {\sqrt{c_\perp^2+g_{\mathrm{amp}}^2s_0^2}}.
\end{aligned}
\label{eq:reshaped-initial-state}
\end{equation}
Here $e^{\ii\zeta_{\mathrm{re}}}$ is a global phase factor.

The finite-$N$ probability of the reshaped initial state in the
selected three-component subspace is
\begin{align}
  p_3^{(\mathrm{re})}
  &=\frac{w_0\,2J(3J+1)/(J+1)^2}
          {1+[4J^2/(J+1)^2-1]w_0}\notag\\
  &\sim6\sqrt{\frac{2}{\pi}}N^{-1/2}.
\end{align}
The normalized marked component satisfies
\begin{equation}
  \frac{\Pi_3\ket{\psi_{\mathrm{i}}^{(\mathrm{re})}}}
       {\sqrt{p_3^{(\mathrm{re})}}}
  =e^{\ii\zeta_{\mathrm{re}}}\ket{\psi_{3,\mathrm{re}}},
  \label{eq:reshaped-marked-component-app}
\end{equation}
so the equal-phase construction gives the exact preparation
relation stated in Sec.~\ref{sec:amplitude-reshaping}. With $p_3^{(\mathrm{dir})}\equiv p_3\sim3\sqrt{2/\pi}N^{-1/2}$, the ratio $p_3^{(\mathrm{re})}/p_3^{(\mathrm{dir})}\to2$ gives $q_{\mathrm{re}}/q_{\mathrm{dir}}\to1/\sqrt2$ and $q_{\mathrm{re}}\simeq0.36\,N^{1/4}$.

\section{Cavity scattering and performance}
\label{app:scattering}

\subsection{Dispersive response and scattering amplitudes}
\label{app:cavity-response}
The numerical calculations use the full frequency-dependent scattering amplitudes, without making the dispersive approximation. The cavity shift follows from Eq.~\eqref{eq:dispersive-hamiltonian}. For the even-$N$ three-component targets, $m=-1,0,+1$ correspond to $n=N/2-1,N/2,N/2+1$. For an ideal one-sided lossless cavity, standard input--output theory~\cite{collett1984,gardiner1985} gives the reflection amplitude~\cite{nagib2025prl,nagib2025pra}
\begin{equation}
    r_n(\omega)
    =
    \frac{
        \ii[\omega-(\omega_0+n\Omega)]-\kappa
    }{
        \ii[\omega-(\omega_0+n\Omega)]+\kappa
    }.
    \label{eq:ideal-reflection}
\end{equation}
Setting $r_n(\bar\omega_p)=e^{\ii\phi}$ gives the photon-frequency condition stated in Sec.~\ref{sec:cavity-phase-operations}. At that frequency, a different component $n'$ has dimensionless detuning $\cot(\phi/2)-(n'-n)d$.

Using the absolute-frequency convention of Sec.~\ref{sec:cavity-phase-operations}, the standard
input--output formalism~\cite{gardiner1985}, and the four-channel
cavity scattering model used in
Refs.~\cite{daiss2019,nagib2025pra}, let
$\kappa=\kappa_r+\kappa_t+\kappa_m$ and define
\begin{equation}
  \mathcal D_n(\omega)
  =
  ng_c^2
  +
  [\ii\Delta_a+\ii(\omega-\omega_0)+\gamma]
  [\ii(\omega-\omega_0)+\kappa].
  \label{eq:scattering-denominator}
\end{equation}
The scattering amplitudes are
\begin{subequations}
\begin{align}
  r_n(\omega)
  &=
  1-
  \frac{
  2\kappa_r[\ii\Delta_a+\ii(\omega-\omega_0)+\gamma]
  }{
  \mathcal D_n(\omega)
  },
  \\
  t_n(\omega)
  &=
  \frac{
  2\sqrt{\kappa_r\kappa_t}
  [\ii\Delta_a+\ii(\omega-\omega_0)+\gamma]
  }{
  \mathcal D_n(\omega)
  },
  \\
  a_n(\omega)
  &=
  \frac{
  2\sqrt{\kappa_r\gamma n g_c^2}
  }{
  \mathcal D_n(\omega)
  },
  \\
  m_n(\omega)
  &=
  \frac{
  2\sqrt{\kappa_r\kappa_m}
  [\ii\Delta_a+\ii(\omega-\omega_0)+\gamma]
  }{
  \mathcal D_n(\omega)
  }.
\end{align}
\end{subequations}
Their environmental labels remain distinct in the Kraus
representation~\cite{kraus1971}, and the reduced atomic channel is obtained by summing
the corresponding Kraus contributions over the distinguishable
output channels.

\subsection{Finite-bandwidth channels and sequence propagation}
\label{app:finite-bandwidth}

We combine the open-cavity input--output model~\cite{collett1984,gardiner1985} with scattering amplitudes that depend on excitation number~\cite{daiss2019,nagib2025pra}. In the operator-sum representation~\cite{kraus1971}, each outgoing channel has a diagonal Kraus operator
\begin{equation}
  K_\mu(\omega)=\sum_{n=0}^{N}\mu_n(\omega)\proj n,
  \label{eq:scattering-kraus-operators-app}
\end{equation}
where the labels $r,t,a,m$ denote reflection, transmission, atomic spontaneous emission, and mirror scattering, respectively. For a normalized spectral envelope $\Phi(\omega)$, the unheralded single-photon channel is
\begin{equation}
    \mathcal E_{\mathrm U}(\rho)
    =
    \int d\omega\,
    \lvert\Phi(\omega)\rvert^2
    \sum_{\mu}
    K_\mu(\omega)\rho K_\mu^\dagger(\omega).
    \label{eq:unheralded-channel}
\end{equation}
The excitation-number dependence of the complex scattering amplitudes reduces atomic coherences when the output modes are traced out. Conditioning on reflection instead gives the trace-decreasing map
\begin{equation}
    \widetilde{\mathcal E}_{\mathrm H}(\rho)
    =
    \int d\omega\,
    \lvert\Phi(\omega)\rvert^2
    K_r(\omega)\rho K_r^\dagger(\omega).
    \label{eq:heralded-channel}
\end{equation}

We use the dimensionless parameters
$C=g_c^2/(\kappa\gamma)$,
$d=\Omega/\kappa=g_c^2/(\Delta_a\kappa)$,
and $w=\sigma/\kappa$ defined in Sec.~\ref{sec:cavity-qed}.
Table~\ref{tab:full-sequence-benchmarks} uses
$\kappa_r=\kappa$, $\kappa_t=\kappa_m=0$,
retains atomic spontaneous emission, and uses a common $d$
for all photon operations within a preparation sequence.

For a requested phase $\phi$ on the target excitation number
$n_{\mathrm{t}}$, the photon center frequency is chosen from the
real-frequency solutions satisfying
\begin{equation}
  \operatorname{Im}\!\left[
  r_{n_{\mathrm{t}}}(\bar\omega_p)e^{-\ii\phi}
  \right]
  =
  0,
  \qquad
  \operatorname{Re}\!\left[
  r_{n_{\mathrm{t}}}(\bar\omega_p)e^{-\ii\phi}
  \right]
  >
  0.
  \label{eq:physical-phase-root}
\end{equation}
When multiple real solutions are present, we select the solution
continuously connected in the dispersive, lossless limit to
the photon-frequency condition in Sec.~\ref{sec:cavity-phase-operations}.
Opposite phases are solved independently at finite loss.
Direct evaluation of the wrapped phase residuals and
excitation-number-resolved phases confirms the target-component
phase conditions for $\pm\alpha$, $\pm\beta$, and $\delta_{\text{re}}$
at the photon center frequencies.

For the Gaussian wave packet used in Sec.~\ref{sec:realistic-performance},
\begin{equation}
  |\Phi(\omega)|^2
  =
  \frac{1}{\sqrt{2\pi}\sigma}
  \exp\!\left[
  -\frac{(\omega-\bar\omega_p)^2}{2\sigma^2}
  \right].
\end{equation}
The channel acts elementwise as
$\rho_{nl}\mapsto\Lambda_{nl}\rho_{nl}$, with
\begin{equation}
  \Lambda_{nl}
  =
  \int d\omega\,
  |\Phi(\omega)|^2
  \sum_\mu
  \mu_n(\omega)\mu_l^*(\omega).
  \label{eq:lambda-matrix}
\end{equation}
For the reflected branch, the channel sum is replaced by
$r_n(\omega)r_l^*(\omega)$.

Starting from the state $\rho_0$ immediately before the first photon operation, the complete sequence is propagated as
\begin{equation}
    \rho_j
    =
    (\mathcal R_j\circ\mathcal E_j)(\rho_{j-1}),
    \label{eq:sequence-propagation}
\end{equation}
where $\mathcal E_j$ is the scattering map and $\mathcal R_j(\rho)=R_j\rho R_j^\dagger$ contains the subsequent global rotations, with $R_j=\mathbb I$ when no rotation follows. The unheralded calculation uses $\mathcal E_{\mathrm U,j}$. For heralded preparation, the same recursion is applied to the unnormalized state using $\widetilde{\mathcal E}_{\mathrm H,j}$.

Full density-matrix rotations are applied between successive diagonal scattering maps. Iterating the heralded map through all $L$ photon operations gives the unnormalized state $\widetilde\rho_L$, with success probability $P_{\mathrm{succ}}=\Tr\widetilde\rho_L$ and conditional output $\rho_{\mathrm H}=\widetilde\rho_L/P_{\mathrm{succ}}$. The final state $\rho_{\mathrm{fin}}$ is $\rho_L$ for unheralded preparation and $\rho_{\mathrm H}$ for heralded preparation.

For the final mixed state $\rho=\rho_{\mathrm{fin}}$, the mean-spin vector and covariance matrix are evaluated as
\begin{subequations}
\label{eq:mixed-moments}
\begin{align}
    \mu_i
    &=
    \Tr(\rho J_i),
    \label{eq:mixed-mean}\\
    \Gamma_{ij}
    &=
    \frac{1}{2}
    \Tr\!\left[
        \rho(J_iJ_j+J_jJ_i)
    \right]
    -
    \mu_i\mu_j.
    \label{eq:mixed-covariance}
\end{align}
\end{subequations}
Let $E$ have orthonormal columns spanning the plane perpendicular to $\bm\mu$. The minimum transverse variance is
\begin{equation}
    (\Delta J_\perp)^2_{\min}
    =
    \lambda_{\min}
    \!\left(
        E^{\mathsf T}\Gamma E
    \right),
    \label{eq:mixed-min-variance}
\end{equation}
and the corresponding Wineland parameter is evaluated as $\xi_R^2=N(\Delta J_\perp)^2_{\min}/\lvert\bm\mu\rvert^2$, following Eq.~\eqref{eq:wineland}.

\begin{figure}[t]
    \centering
    \includegraphics[width=\columnwidth]{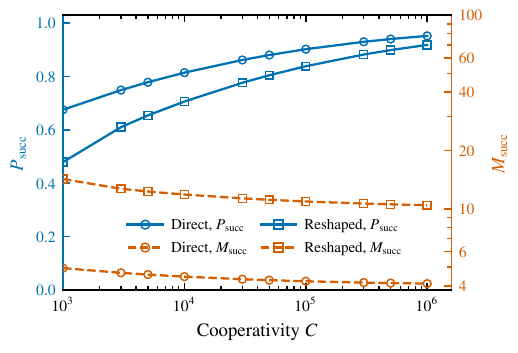}
    \caption{Heralding success probability $P_{\mathrm{succ}}$ (left, linear axis) and mean incident-photon consumption per successful preparation $M_{\mathrm{succ}}$ (right, logarithmic axis) versus cavity cooperativity $C$ for $N=50$ at $w=0.1$.}
    \label{fig:heralding-tradeoff}
\end{figure}

\subsection{Heralding success probability and photon cost}
\label{app:heralding-cost}

For heralded preparation, let $S_\ell$ be the survival probability after the first $\ell$ successful photon heralds, with $S_0=1$ and $S_L=P_{\mathrm{succ}}$. Under an early-abort-and-restart strategy, the mean number of incident photons consumed per successful preparation is $M_{\mathrm{succ}}=\sum_{\ell=0}^{L-1}S_\ell/P_{\mathrm{succ}}$. Unlike the fixed photon count $N_{\mathrm{ph}}$ required to execute a complete protocol sequence in Sec.~\ref{sec:physical-resources}, $M_{\mathrm{succ}}$ accounts for early termination after a failed herald and the subsequent restart, and is therefore generally not equal to $N_{\mathrm{ph}}/P_{\mathrm{succ}}$.

For $N=50$, Fig.~\ref{fig:heralding-tradeoff} shows that the direct-projection protocol has a higher heralding success probability and a lower mean incident-photon consumption per successful preparation over the displayed cooperativity range. The corresponding heralded target-state fidelities are shown in the inset of Fig.~\ref{fig:realistic-performance}(b); both increase with cooperativity over the displayed range and reach values close to unity at the largest cooperativity shown. Together with the squeezing results in Sec.~\ref{sec:realistic-performance}, this comparison illustrates the trade-off between squeezing performance and implementation cost.

\subsection{Heuristic error scaling}
\label{app:error-estimates}

We use the $C$- and $d$-dependence derived in
Ref.~\cite{nagib2025pra} for Dicke-selective phase inversions as a
heuristic guide for the non-$\pi$ phase operations appearing in the
present protocol.
For a nonzero target excitation $n_t$, the single-operation errors are
written as
\begin{subequations}
\begin{align}
  \epsilon_{\mathrm U}
  &\sim
  \frac{A(\phi)}{d^2}
  +
  B(\phi)\frac{n_td^2}{C}
  +
  O(w^2),
  \\
  \epsilon_{\mathrm H}
  &\sim
  \frac{A'(\phi)}{d^2}
  +
  B'(\phi)\frac{n_t^2d^4}{C^2}
  +
  O(w^2).
\end{align}
\end{subequations}
The phase dependence is absorbed into
$A(\phi)$, $B(\phi)$, $A'(\phi)$, and $B'(\phi)$.
Assuming these coefficients remain of order unity, optimization of
the displayed terms gives
$d_{\mathrm{opt}}\sim(C/n_t)^{1/4}$ and
$\epsilon_{\mathrm U}^{\mathrm{opt}}\sim\sqrt{n_t/C}+O(w^2)$
for the unheralded case, and
$d_{\mathrm{opt}}\sim(C/n_t)^{1/3}$ and
$\epsilon_{\mathrm H}^{\mathrm{opt}}\sim(n_t/C)^{2/3}+O(w^2)$
for the heralded case.

For the Dicke-selective operations acting on the central components
of the present protocols,
$n_t=J+O(1)\simeq N/2$.
With $L=O(N^{1/4})$ such photon operations and approximately linear
accumulation of their cooperativity-dependent single-operation errors,
\begin{equation}
  E_{\mathrm U}^{(\mathrm{lin})}
  \sim
  \frac{N^{3/4}}{\sqrt C},
  \qquad
  E_{\mathrm H}^{(\mathrm{lin})}
  \sim
  \frac{N^{11/12}}{C^{2/3}}.
\end{equation}
Keeping these contributions bounded suggests
$C\gtrsim N^{3/2}$ for the unheralded case and
$C\gtrsim N^{11/8}$ for the heralded case.

\bibliography{references}

\end{document}